\input{aipcheck}

\documentclass[
    ,final            
  ]
  {aipproc}

\layoutstyle{6x9}

\begin{document}

\title{$P_{11}$ Resonance Extracted from $\pi N$ Data and Its Stability}

\classification{14.20.Gk, 13.75.Gx, 13.60.Le}
\keywords      {$\pi N$ scattering, amplitude analysis, Roper resonance}

\author{Satoshi X. Nakamura}{
  address={Excited Baryon Analysis Center (EBAC)\\ 
Thomas Jefferson National Accelerator Facility, Newport News, Virginia 23606, USA}
}

\begin{abstract}
 We study the stability of resonance poles in $\pi N$ $P_{11}$ 
 partial wave, particularly the Roper resonance, 
 by varying parameters significantly within the EBAC dynamical
 coupled-channels model, keeping a good fit to the empirical amplitude. 
 We find that two Roper poles are stable against the
 variation. However, for higher energies, the number of poles can change
 depending on how the parameters are fitted  within error bars.
 We also developed a model
 with a bare nucleon which forms the physical nucleon by being
 dressed by the meson-cloud. 
We still find a good stability of the Roper poles. 
\end{abstract}

\maketitle


\section{Introduction}

For extracting $N^*$ information, first, one needs to construct
a reaction model through a comprehensive analysis of data.
Then, pole positions and vertex form factors are extracted from the
model with the use of the analytic continuation.
Therefore, the $N^*$ information extracted in this manner is inevitably
model-dependent. 
Thus, commonly asked
questions are how model-dependent are the extracted resonance
parameters, and how precise do data have to be for stable resonance extraction ?
These are the questions we address at
Excited Baryon Analysis Center (EBAC) at JLab\cite{hnls10},
within a dynamical coupled-channels model (EBAC-DCC)~\cite{msl07}. 
We focus on the $\pi N$ $P_{11}$ partial wave and 
the stability of its pole positions, particularly those corresponding to the
Roper resonance.
In the region near Roper $N(1440)$, two poles
close to the $\pi\Delta$ threshold were found in 
our recent extraction~\cite{sjklms10} from the JLMS model~\cite{jlms07} 
(JLMS is one of EBAC-DCC model),
while only one pole in the
similar energy region was reported in some other analyses.
We examine the stability of this two-pole structure against the
following variation, keeping a good reproduction of 
SAID single-energy (SAID-SES) solution~\cite{said-1}.
\begin{itemize}
 \item Large variation of the parameters of the meson-baryon
        and bare $N^*$ parameters of the EBAC-DCC model.
 \item Inclusion of a bare nucleon state:
The analytic structure
of this model is different from the EBAC-DCC model
in the region near the nucleon pole~\cite{jklmss09},  .
\end{itemize}

\section{Dynamical coupled-channels models}

Here, we briefly describe
the EBAC-DCC model and the bare nucleon model.
The EBAC-DCC model contains $\pi N$, $\eta N$ and $\pi\pi N$ channels
and the $\pi\pi N$ channel has $\pi\Delta$, $\rho N$ and $\sigma N$
components.
These meson-baryon (MB) channels are connected with each other by meson-baryon
interactions ($v_{MB,M'B'}$), or excited to bare $N^*$ states by vertex
interactions ($\Gamma_{MB\leftrightarrow N^*}$).
With these interactions, 
The partial-wave amplitude for the
$M(\vec{k})+B(-\vec{k}) \to M'(\vec{k}')+B'(-\vec{k}')$
reaction can be conveniently decomposed into two parts as
$T_{MB,M'B'}(k,k',E)=t_{MB,M'B'}(k,k',E)+t^{R}_{MB,M'B'}(k,k',E)$.
The first term is obtained by solving the coupled-channels
Lippmann-Schwinger equation with $v_{MB,M'B'}$ only.
The second term is
associated with the bare $N^*$ states, and given by
\begin{eqnarray}
t^{R}_{MB,M^\prime B^\prime}(k,k',E)&=& \sum_{i,j}
\bar{\Gamma}_{MB \to N^*_i}(k,E) [D(E)]_{i,j}
\bar{\Gamma}_{N^*_j \to M^\prime B^\prime}(k',E),
\label{eq:tmbmb-r}
\end{eqnarray}
where the dressed vertex function 
$\bar{\Gamma}_{N^*_j \to M^\prime B^\prime}(k,E)$ is 
 calculated by convoluting
the bare vertex ${\Gamma}_{N^*_j \to M^\prime B^\prime}(k)$ with
the amplitudes $t_{MB,M^\prime B^\prime}(k,k',E)$.
The inverse of the propagator of dressed $N^*$ states in
Eq.~(\ref{eq:tmbmb-r})
is \begin{equation}
[D^{-1}(E)]_{i,j} = (E - m^0_{N^*_i})\delta_{i,j} - \Sigma_{i,j}(E) ,
\label{eq:nstar-selfe}
\end{equation}
where $m^0_{N^*_i}$  is the bare mass of the $i$-th $N^*$ state,
and the $N^*$ self-energy is defined by
\begin{equation}
\Sigma_{i,j}(E)= \sum_{MB} \int_{C_{MB}}  q^2 dq 
\bar{\Gamma}_{N^*_j \to M B}(q,E) G_{MB}(q,E) {\Gamma}_{MB \to
N^*_i}(q,E) \ ,
\label{eq:nstar-g}
\end{equation}
where $G_{MB}$ is the meson-baryon propagator, and
$C_{MB}$ is the integration contour in the complex-$q$ plane used
for the channel $MB$.

To examine further the model dependence of resonance extractions, 
it is useful to also  perform
analysis using models with a bare nucleon, as developed in
Ref.~\cite{afnan}.
Within the formulation of EBAC-DCC model,
such a model can be obtained by
adding a bare nucleon ($N_0$) state with mass $m^0_N$
and $N_0\rightarrow MB $ vertices and
removing the direct $MB \rightarrow N \rightarrow M'B'$
in the meson-baryon
interactions $v_{MB,M'B'}$.
All numerical procedures for this model
are identical to that used for the EBAC-DCC model,
except that the resulting amplitude must satisfy the nucleon pole
condition:
\begin{equation}
t^R_{\pi N,\pi N}(k\to k_{\rm{on}},k\to k_{\rm{on}},E\rightarrow m_N ) 
= -\frac{[F_{\pi NN} (k_{\rm{on}})]^2}{E - m^0_N - \tilde{\Sigma}(m_N)} .
\label{eq:pole-t}
\end{equation}
with 
\begin{eqnarray}
m_N= m^0_N + \tilde{\Sigma}(m_N)
\qquad {\rm and} \qquad
F_{\pi NN}(k_{\rm{on}})= F^{\rm{phys.}}_{\pi NN}(k_{\rm{on}}) \ .
\label{eq:pole-m}
\end{eqnarray}
Here we have used the on-shell momentum defined by
$E=\sqrt{m_N^2+k^2_{\rm{on}}}+\sqrt{m_\pi^2+k^2_{\rm{on}}}$.
Also, $\tilde{\Sigma}(m_N)$ is
the self-energy for the nucleon.
More details for the calculational 
procedure following Afnan and Pearce
is found in Refs.~\cite{hnls10,afnan}.

\section{Results}

Now we show our numerical results to examine the stability of the
$P_{11}$ poles. 
We present results from various fits by
varying the dynamical content of the EBAC-DCC model,
and by using a model with a bare nucleon.
We show in figures the quality of fits of these models, and 
in Table \ref{tab:p11-tab1} the pole positions from the models
as well as $\chi^2$ per data point ($\chi^2_{pd}$).
We find the poles with the
method of analytic continuation discussed in detail
in Refs.~\cite{sjklms10,ssl09}.
In Table \ref{tab:p11-tab1}, we also present pole positions from 
JLMS\cite{jlms07} and SAID-EDS (energy-dependent)\cite{said-1}.

\begin{table}[b]
\renewcommand{\arraystretch}{1.3}
\tabcolsep=3.0mm
\begin{tabular}{ccccccc}\hline
Model           & $upuupp$   & $upuppp$  & $uuuupp$  & $uuuuup$& $\chi^2_{pd}$   \\ 
\hline
SAID-EDS        & (1359, 81) & (1388, 83) &    ---    &  ---        & 2.94 \\
JLMS            & (1357, 76) & (1364, 105)&    ---    & (1820, 248) & 3.55 \\
2$N^*$-3p       & (1368, 82) & (1375, 110)&    ---    & (1810, 82)  & 3.28 \\
2$N^*$-4p       & (1372, 80) & (1385, 114)& (1636, 67)& (1960, 215) & 3.36 \\
\hline
1$N_0$1$N^*$-3p & (1363, 81) & (1377, 128)&    ---    & (1764, 137) & 2.51 \\\hline
\end{tabular}
\caption{\label{tab:p11-tab1}
The resonance pole positions $M_R$ for $P_{11}$
[listed as ($\rm{Re}M_R$, $-\rm{Im} M_R$) in the unit of MeV] extracted from
 various parameter sets.
The location of the pole is specified by, e.g.,
$(s_{\pi N},s_{\eta N},s_{\pi\pi N},s_{\pi\Delta},s_{\rho N},s_{\sigma N})=(upuupp)$,
where $p$ and $u$ denote the physical and unphysical sheets for a
given reaction channel, respectively. $\chi^2_{pd}$ is $\chi^2$ per data point.}
\end{table}

\begin{figure}[t]
\begin{minipage}[t]{75mm}
   \includegraphics[width=75mm]{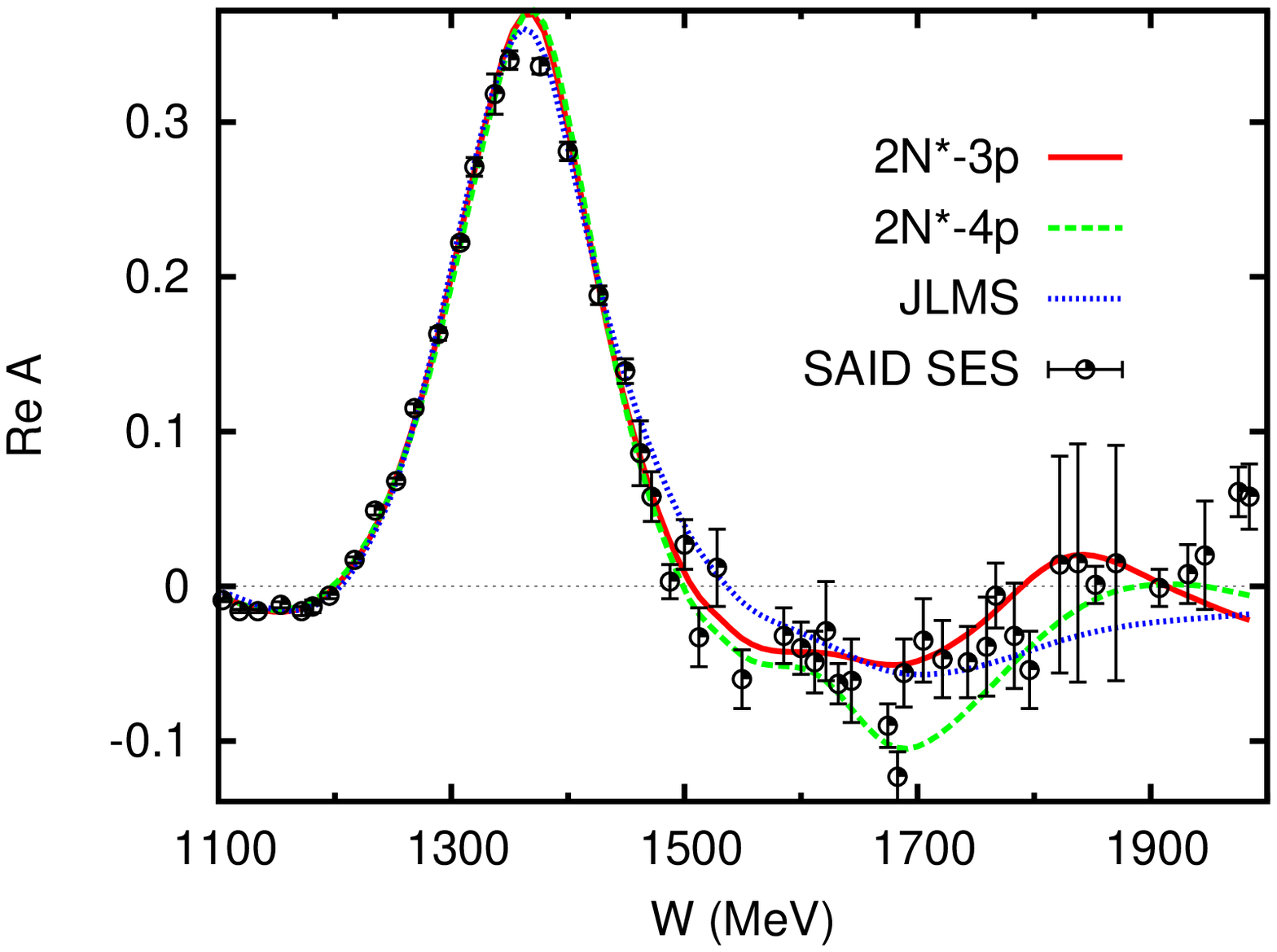}
 \end{minipage}
\begin{minipage}[t]{75mm}
   \includegraphics[width=75mm]{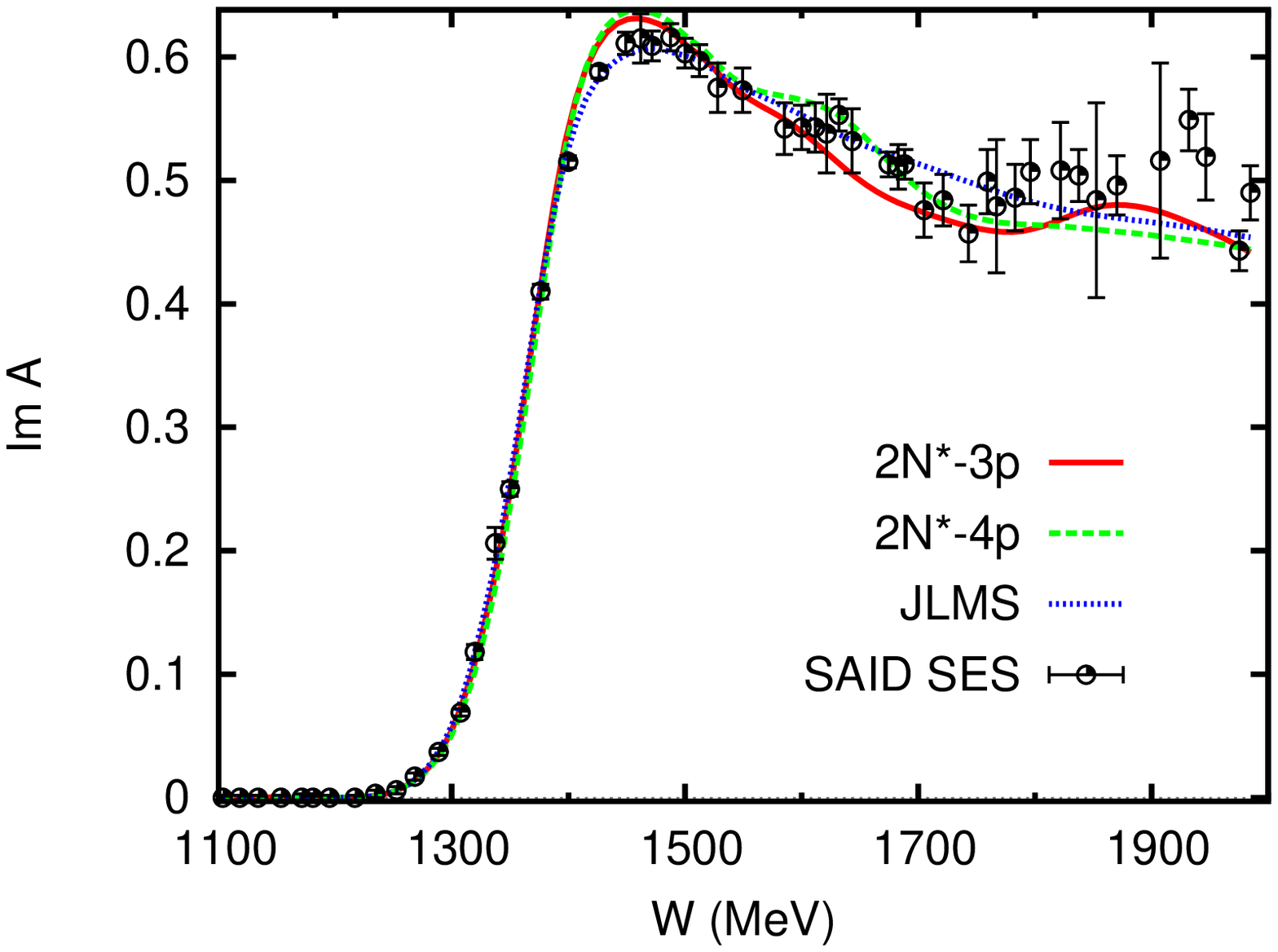}
 \end{minipage}
\caption{\label{fig:p11-fig2}
The real (left) and imaginary (right) parts of the on-shell $P_{11}$
amplitudes as a function of the $\pi N$ invariant mass $W$ (MeV).
$A$ is unitless in the convention of Ref.~\cite{said-1}.
}
\end{figure}

First
we varied both the parameters for the meson-baryon interactions
($v_{MB,M'B'}$) and parameters associated with bare $N^*$ states 
($m_{N^*}^0$, $\Gamma_{N^*\leftrightarrow MB}$) within EBAC-DCC model.
The obtained meson-baryon interactions are quite different from those of
JLMS.
We obtained several fits which are different in how the oscillatory
behavior of SAID-SES amplitude for higher $W$ is fitted.
The results from the 2$N^*$-3p (dotted curves) and 2$N^*$-4p (dashed curves)
fits are compared with the JLMS fit (solid curves) 
in Fig.~\ref{fig:p11-fig2}.
The resulting resonance poles
are listed in the 3th and 4th rows of Table~\ref{tab:p11-tab1}.
Here we see the first two poles near the $\pi\Delta$ threshold from
both fits agree well with the poles from the JLMS fit. This seems to further support the
conjecture that these two poles are mainly sensitive to the
data  below $W\sim 1.5$ GeV where the SAID-SES has rather small errors.
However, the 2$N^*$-4p fit has one more pole at 
$M_R= 1630 -i45$ MeV. This
is perhaps related to its oscillating structure near $W\sim 1.6$ GeV
(dashed curves), as shown in the Figs.~\ref{fig:p11-fig2}.
On the other hand, this 
resonance pole could be
fictitious since the fit 2$N^*$-3p (dotted curve) with only three poles
are equally acceptable within the fluctuating experimental errors.
Our result suggests that it is important to have more accurate data
in the high $W$ region for a high precision resonance extraction.

Next, we show our results obtained with the bare nucleon model, and then
address the question whether difference in the analytic structure of the
$\pi N$ amplitude below $\pi N$ threshold strongly affects the resonance
extractions. 
The bare nucleon model is fitted to SAID-SES, and at the same time, to
the nucleon pole conditions Eq.~(\ref{eq:pole-m}).
Meanwhile, the original EBAC-DCC model has different singular structure
below the $\pi N$ threshold.
The question is whether such differences 
can lead to very different resonance poles.
Our fit of the bare nucleon model agree very well with JLMS below
$W=1.5$ GeV, while
their differences are significant in the high $W$ region.
The corresponding resonance poles are given in Table~\ref{tab:p11-tab1}.
We also see here that the first two
poles near the $\pi\Delta$ threshold are close to those of JLMS.
Our results seem to indicate that these two poles are rather insensitive to
the analytic structure of the amplitude in the region below $\pi N$ threshold, 
and are mainly determined by the data in the region 
$ m_N+m_\pi\leq W \leq 1.6 $ GeV.


\begin{theacknowledgments}
The author would like to thank H. Kamano, T.-S. H. Lee and
T. Sato for their collaborations at EBAC.
This work is supported by the U.S. Department of Energy, Office of
 Nuclear Physics Division, under Contract No. DE-AC05-06OR23177
under which Jefferson Science Associates operates Jefferson Lab.
\end{theacknowledgments}



\bibliographystyle{aipproc}   

\begin{thebibliography}{99}

\bibitem{hnls10}
H.~Kamano,  S.~X. Nakamura, T.-S.~H. Lee, and T.~Sato,
\emph{Phys. Rev. C} \textbf{81}, 065207 (2010).

\bibitem{msl07}
A.~Matsuyama, T.~Sato, and T.-S.~H. Lee,
\emph{Phys.\ Rep.} \textbf{439}, 193 (2007).

\bibitem{sjklms10}
N.~Suzuki, B.~Juli\'a-D\'iaz, H.~Kamano, T.-S.~H. Lee, A.~Matsuyama, 
and T.~Sato,
\emph{Phys.\ Rev.\ Lett.} \textbf{104}, 042302 (2010).

\bibitem{jlms07}
B.~Juli\'a-D\'iaz, T.-S.~H. Lee, A.~Matsuyama, and T.~Sato,
 \emph{Phys. Rev. C} \textbf{76}, 065201 (2007).

\bibitem{said-1}
R.~A.~Arndt, W.~J. Briscoe, I.~I. Strakovsky, and R.~L. Workman,
\emph{Phys.\ Rev.\ C} \textbf{74}, 045205 (2006).

\bibitem{jklmss09}
B.~Juli\'a-D\'iaz, H.~Kamano, T.-S.~H. Lee, A.~Matsuyama, T.~Sato, and N.~Suzuki,
\emph{Chin.\ J.\ Phys.}\ \textbf{47}, 142 (2009).

\bibitem{cw90}
R.~E. Cutkosky, and S.~Wang,
\emph{Phys.\ Rev.\ D} \textbf{42}, 235 (1990).

\bibitem{afnan}
B.~C. Pearce, and I.~R. Afnan,
\emph{Phys.\ Rev.\ C} \textbf{34}, 991 (1986);
\emph{Phys.\ Rev.\ C} \textbf{40}, 220 (1989).

\bibitem{ssl09}
N.~Suzuki, T.~Sato, and T.-S.~H. Lee,
\emph{Phys.\ Rev.\ C} \textbf{79}, 025205 (2009); arXiv:1006.2196[nucl-th].



\end{thebibliography}

\end{document}